\documentclass[%
 reprint,
superscriptaddress,
 amsmath,amssymb,
 aps,
]{revtex4-2}

\usepackage{graphicx}
\usepackage{dcolumn}
\usepackage{bm}
\usepackage{hyperref}
\usepackage{chngcntr}
\usepackage{color}

\begin{document}


\title{Observation of Colossal Terahertz Magnetoresistance and Magnetocapacitance in a Perovskite Manganite}

\author{Fuyang~Tay}
\affiliation{Department of Electrical and Computer Engineering, Rice University, Houston, Texas 77005, USA}
\affiliation{Applied Physics Graduate Program, Smalley--Curl Institute, Rice University, Houston, Texas 77005, USA}

\author{Swati~Chaudhary}
\affiliation{Department of Physics, University of Texas at Austin, Austin, Texas 78712, USA}
\affiliation{Department of Physics, Northeastern University, Boston, Massachusetts 02115, USA}
\affiliation{Department of Physics, Massachusetts Institute of Technology, Cambridge, Massachusetts 02139, USA}

\author{Jiaming~He}
\affiliation{Materials Science and Engineering Program, Mechanical Engineering, University of Texas at Austin, Austin, Texas 78712, USA}

\author{Nicolas~Marquez~Peraca}
\affiliation{Department of Physics and Astronomy, Rice University, Houston, Texas 77005, USA}

\author{Andrey~Baydin}
\affiliation{Department of Electrical and Computer Engineering, Rice University, Houston, Texas 77005, USA}%
\affiliation{Smalley--Curl Institute, Rice University, Houston, Texas 77005, USA}

\author{Gregory~A.~Fiete}
\affiliation{Department of Physics, Northeastern University, Boston, Massachusetts 02115, USA}
\affiliation{Department of Physics, Massachusetts Institute of Technology, Cambridge, Massachusetts 02139, USA}

\author{Jianshi~Zhou}
\affiliation{Materials Science and Engineering Program, Mechanical Engineering, University of Texas at Austin, Austin, Texas 78712, USA}

\author{Junichiro~Kono}
\email{kono@rice.edu}
\affiliation{Department of Electrical and Computer Engineering, Rice University, Houston, Texas 77005, USA}
\affiliation{Department of Physics and Astronomy, Rice University, Houston, Texas 77005, USA}
\affiliation{Smalley--Curl Institute, Rice University, Houston, Texas 77005, USA}
\affiliation{Department of Materials Science and NanoEngineering, Rice University, Houston, Texas 77005, USA}

\date{\today}

\begin{abstract}
We have studied the terahertz response of a bulk single crystal of La$_{0.875}$Sr$_{0.125}$MnO$_3$ at around its Curie temperature, observing large changes in the real and imaginary parts of the optical conductivity as a function of magnetic field.  The terahertz resistance and capacitance extracted from the optical conductivity rapidly increased with increasing magnetic field and did not show any sign of saturation up to 6\,T, reaching 60\% and 15\%, respectively, at 180\,K.  The observed terahertz colossal magnetoresistance and magnetocapacitance effects can be qualitatively explained by using a two-component model that assumes the coexistence of two phases with vastly different conductivities. These results demonstrate the potential use of perovskite manganites for developing efficient terahertz devices based on magnetic modulations of the amplitude and phase of terahertz waves.
\end{abstract}

\maketitle


\section{Introduction}
Probing magnetotransport properties in the terahertz (THz) frequency range can provide fundamental insights into the states and dynamics of charge and spin carriers in materials while it can also lead to new device concepts for applications in THz technology~\cite{Tonouchi2007NP,Mittleman2017JAP}. Measurements of the amplitude and phase of the transmitted or reflected THz electric fields allow one to determine multiple transport parameters simultaneously from the real and imaginary parts of the optical conductivity without any metallic contacts~\cite{Lloyd-HughesJeon2012JIMTW,JinEtAl2015NP,SpiesEtAl2020JPCC}. For instance, direct current (DC) conductivities and scattering times for the majority- and minority-spin electrons involved in the giant magnetoresistance (MR) effect have been determined unambiguously from THz measurements~\cite{JinEtAl2015NP}. 
Also, the magnetic tunability of optical constants in the THz regime may find uses in THz device development, such as magnetic THz modulators~\cite{RahmEtAl2013JIMTW,MaEtAl2019R}. Further, a maximal modulation depth of 60\% in the THz frequency range has been demonstrated in a magnetic tunnel junction using a low magnetic field of 30\,mT~\cite{JinEtAl2020PRA}. 

Recently, colossal magnetoresistance (CMR) in La$_{0.7}$Sr$_{0.3}$MnO$_3$ thin films has been shown to persist up to THz frequencies~\cite{Lloyd-HughesEtAl2017NL}. The perovskite manganite (La$_{1-x}$Sr$_{x}$MnO$_3$ or LSMO) is a complex many-body system in which multiple competing degrees of freedom lead to a variety of phase transitions as a function of Sr composition $x$, temperature ($T$), and magnetic field ($H$)~\cite{Ramirez1997JPCM,DagottoEtAl2001PR,ZhouGoodenough2015PRB}. At $x=0.3$, LSMO is metallic at room temperature, exhibiting a Drude-like conductivity spectrum, i.e., the real part of the complex optical conductivity, $\sigma_1(\omega)$, decreases with increasing frequency, $\omega$. Magnetic-field-dependent THz measurements demonstrated that the $\sigma_1(\omega)$ of La$_{0.7}$Sr$_{0.3}$MnO$_3$ increases substantially with magnetic field~\cite{Lloyd-HughesEtAl2017NL}. However, the THz optical constants of dielectric CMR manganites in an external magnetic field have not been investigated, and the question of whether the colossal magnetocapacitance (MC) (or magnetodielectric) effect observed at kHz frequencies~\cite{MaminEtAl2007PRB} would persist at THz frequencies remains unanswered.

Here we use THz time-domain magnetospectroscopy (THz-TDMS) to extract the complex optical constants of a bulk single crystal of LSMO with $x=0.125$ in the THz frequency range as a function of $\omega$, $T$, and $H$. In contrast to La$_{0.7}$Sr$_{0.3}$MnO$_3$~\cite{Lloyd-HughesEtAl2017NL}, La$_{0.875}$Sr$_{0.125}$MnO$_3$ exhibits relatively low conductivity~\cite{ZhouGoodenough2015PRB} and non-Drude behavior in the conductivity spectrum~\cite{PimenovEtAl1999PRB}, which implies that the conductivity mechanism in the material is dominated by electron hopping or tunneling between localized states. A previous study indicated that a transition from electron hopping to Drude conduction occurs at $x\sim0.14$~\cite{ZhouGoodenough2000PRB}. Our results show the coexistence of colossal THz MR and MC in La$_{0.875}$Sr$_{0.125}$MnO$_3$ near the Curie temperature, $T_\text{C}$. Both the amplitude and phase of transmitted THz radiation were significantly modulated by an externally applied magnetic field. The colossal THz MR and MC effects did not show any saturation within the range of magnetic fields in our measurements, reaching 60\% and 15\%, respectively, at $\omega/2\pi=0.4$\,THz, $T=180$\,K, and $\mu_0H=6$\,T. The $H$-induced permittivity (relative) changes exceeded 4. The $H$-induced conductivity changes at DC and THz frequencies were found to be almost identical. The spectral shapes of the real part of the conductivity ($\sigma_1$) and permittivity ($\varepsilon_1$) did not change significantly with increasing $H$. We used a two-component model accounting for two phases with different conductivities~\cite{EfrosShklovskii1976PSSB} to explain the simultaneous increase of $\sigma_1$ and $\varepsilon_1$ with decreasing $T$ or increasing $H$. The colossal THz MR and MC in mixed-valence manganites may lead to applications in THz devices such as spintronics~\cite{WalowskiMunzenberg2016JoAP} and magnetically driven THz amplitude and phase modulators~\cite{RahmEtAl2013JIMTW,MaEtAl2019R}.

\section{Sample and Experimental Methods}
\label{sec:methods}
We studied a $c$-axis-oriented single crystal of La$_{0.875}$Sr$_{0.125}$MnO$_3$, grown by using the floating zone method with an image furnace (NEC M-35HD)~\cite{ZhouGoodenough2015PRB}. A stoichiometric mixture of La$_2$O$_3$, SrCo$_3$, and Mn$_2$O$_3$ was used for the feed and seed rods. The dimensions of the crystal were 0.31\,$\times$\,0.72\,$\times$\,0.88\,mm$^{3}$ with the $c$-axis along the shortest dimension of the pellet. 
We obtained basic information on DC transport properties of La$_{0.875}$Sr$_{0.125}$MnO$_3$ through $T$- and $H$-dependent conductivity measurements on a different crystal using a Physical Property Measurement System (Quantum Design, Inc.). The temperature range was between 70\,K and 320\,K.

We determined the complex optical constants of the La$_{0.875}$Sr$_{0.125}$MnO$_3$ crystal in the THz frequency range by performing THz-TDMS measurements~\cite{WangEtAl2007OL,WangEtAl2010OE,ArikawaEtAl2011PRB,ArikawaEtAl2012OE,ZhangEtAl2014PRLa,ZhangEtAl2016NPb,LiEtAl2018NP,LiEtAl2018S,LiEtAl2019PRB,BaydinEtAl2020FO,BaydinEtAl2022PRLa}. The near-infrared (775\,nm) output beam of a Ti:sapphire regenerative amplifier (1\,kHz repetition rate, 200\,fs pulse duration, Clark-MXR, Inc., CPA-2001) was split into a pump beam and a probe beam. The pump beam generated linearly polarized broadband THz radiation in a ZnTe crystal through optical rectification. The generated THz beam was focused onto the sample mounted on the sample holder of a 10\,T superconducting magnet cryostat (Spectromag, Oxford Instruments, Inc.). An external static magnetic field was applied parallel to the THz beam path, which is normal to the sample surface (Faraday configuration); see Fig.~\ref{fig:time}(a). To minimize birefringence effects in the crystal, we set the polarization of the THz beam to be along the $a$-axis; see Supplement 1. We measured the time-domain waveform of the transmitted THz radiation, $E_\text{sample}(t)$, with another ZnTe crystal via electro-optic sampling with controlled time delays. A reference signal, $E_\text{reference}(t)$, was obtained by repeating the measurements in the absence of the sample.

\section{Results and Discussion}
\begin{figure}[hbt!]
    \centering
    \includegraphics[width=0.45\textwidth]{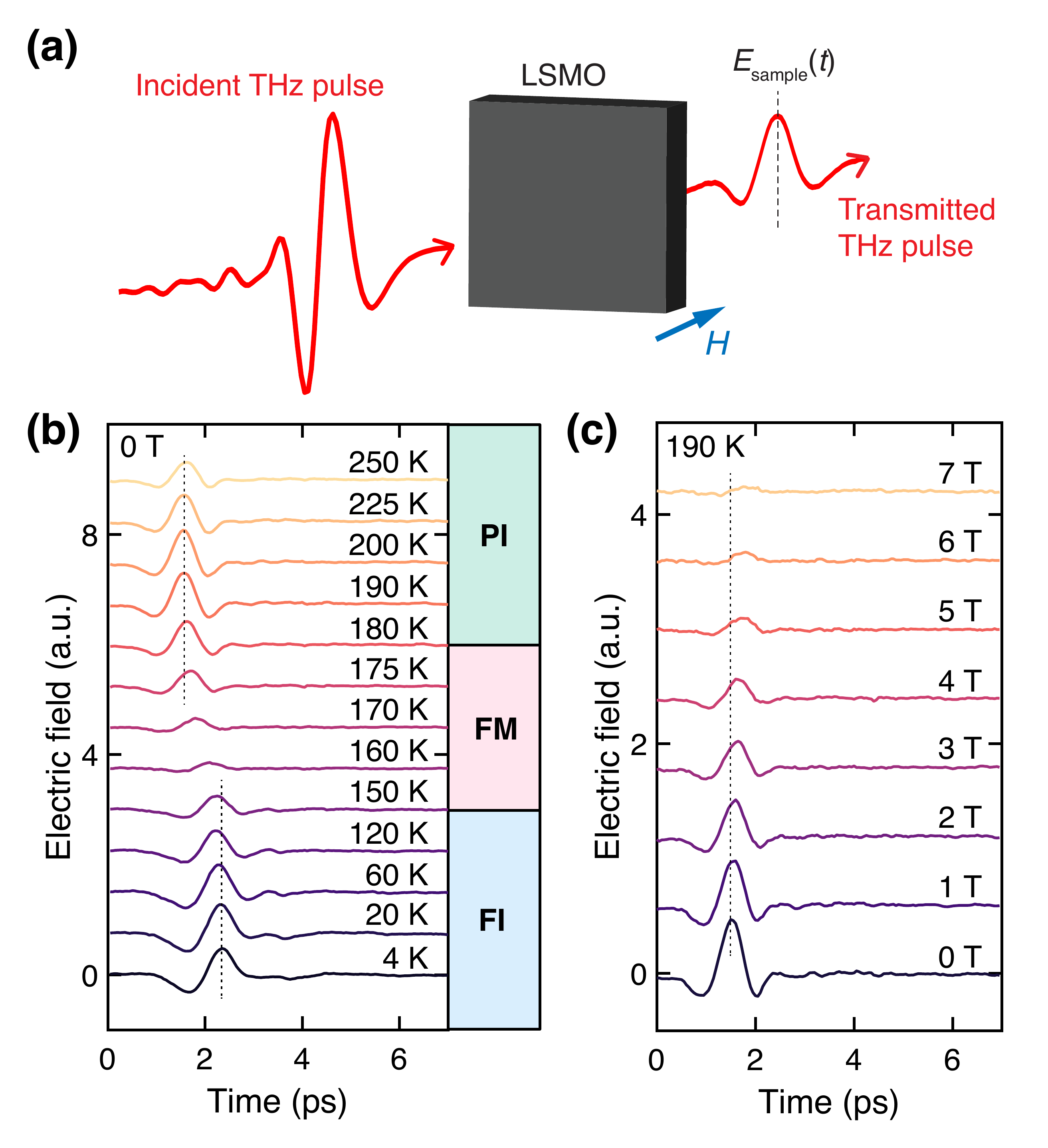}
    \caption{Experimental setup and the waveform of the transmitted THz pulse in the time domain at various temperatures and magnetic fields. (a)~The THz pulse was normally incident upon the sample surface, propagating along the $c$-axis of the La$_{0.875}$Sr$_{0.125}$MnO$_3$ sample. The magnetic field was applied parallel to the light path (Faraday geometry). The transmitted THz waveforms in the time domain are plotted at different (b)~temperatures and (c)~magnetic fields. PI, FM, and FI in (b) denote the different magnetic and electronic phases of the sample, representing the paramagnetic (P) and insulating (I), ferromagnetic (F) and metallic (M), and ferromagnetic (F) and insulating (I) phases, respectively. The vertical dashed lines highlight the phase-dependent time delay due to changes in the real part of the refractive index.  }
    \label{fig:time}
\end{figure}

\begin{figure}[hbt!]
    \centering
    \includegraphics[width=0.45\textwidth]{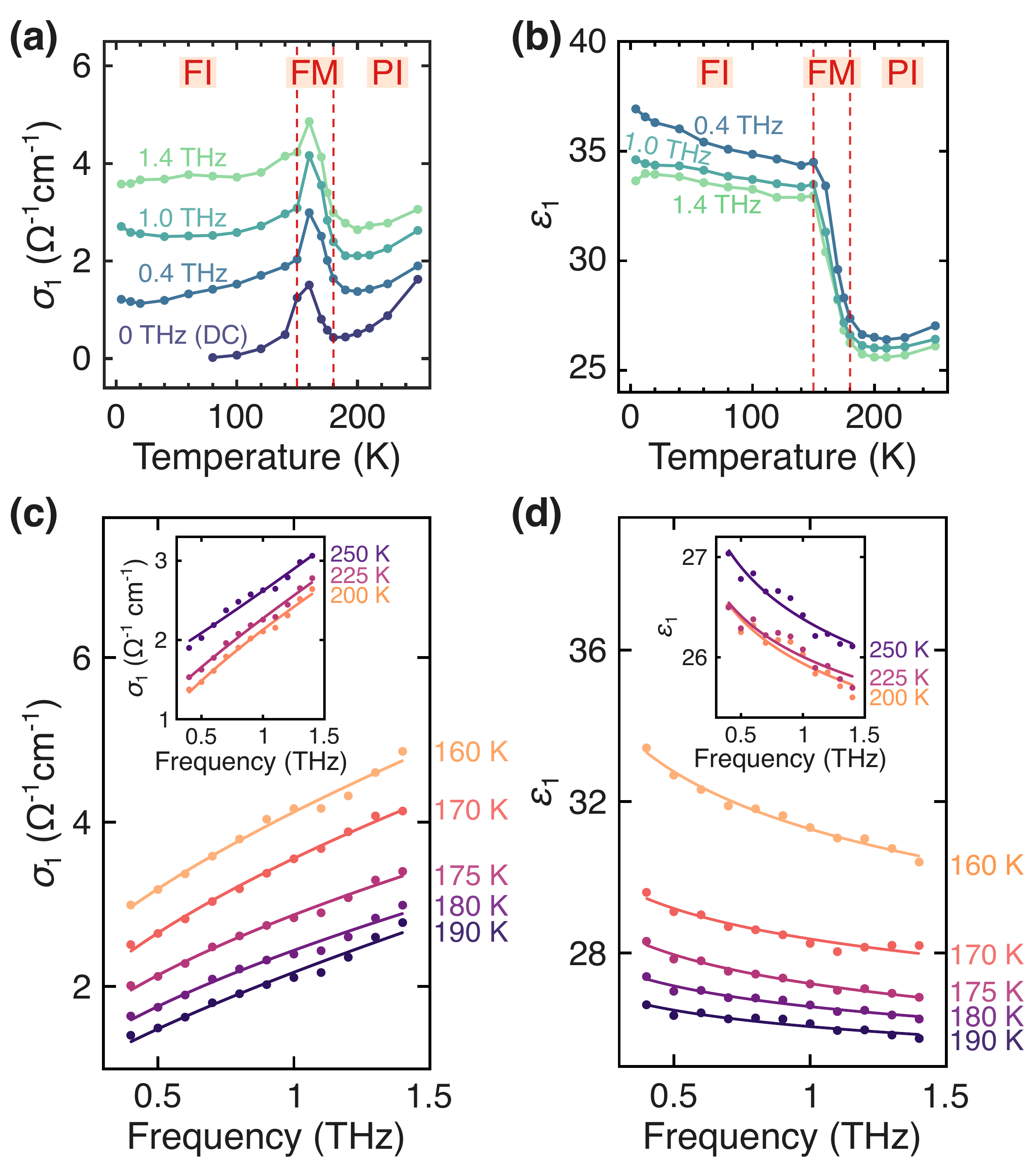}
    \caption{Temperature dependence of the real part of the conductivity, $\sigma_1(\omega)$, and the real part of the permittivity, $\varepsilon_1(\omega)$. Temperature dependence of (a) $\sigma_1$ and (b) $\varepsilon_1$ at different frequencies ($\omega$). The red dashed lines represent the phase transitions that occur at $T_\text{C}$$\sim$180\,K (PI to FM) and $T_\text{CO}$$\sim$150\,K (FM to FI). (c)~$\sigma_1(\omega)$, and (d)~$\varepsilon_1(\omega)$, at $T$ between 160\,K and 190\,K. The insets show $\sigma_1(\omega)$ and $\varepsilon_1(\omega)$ at $T$ above 190\,K.}
    \label{fig:Tdep}
\end{figure}

\begin{figure}[hbt!]
    \centering
    \includegraphics[width=0.45\textwidth]{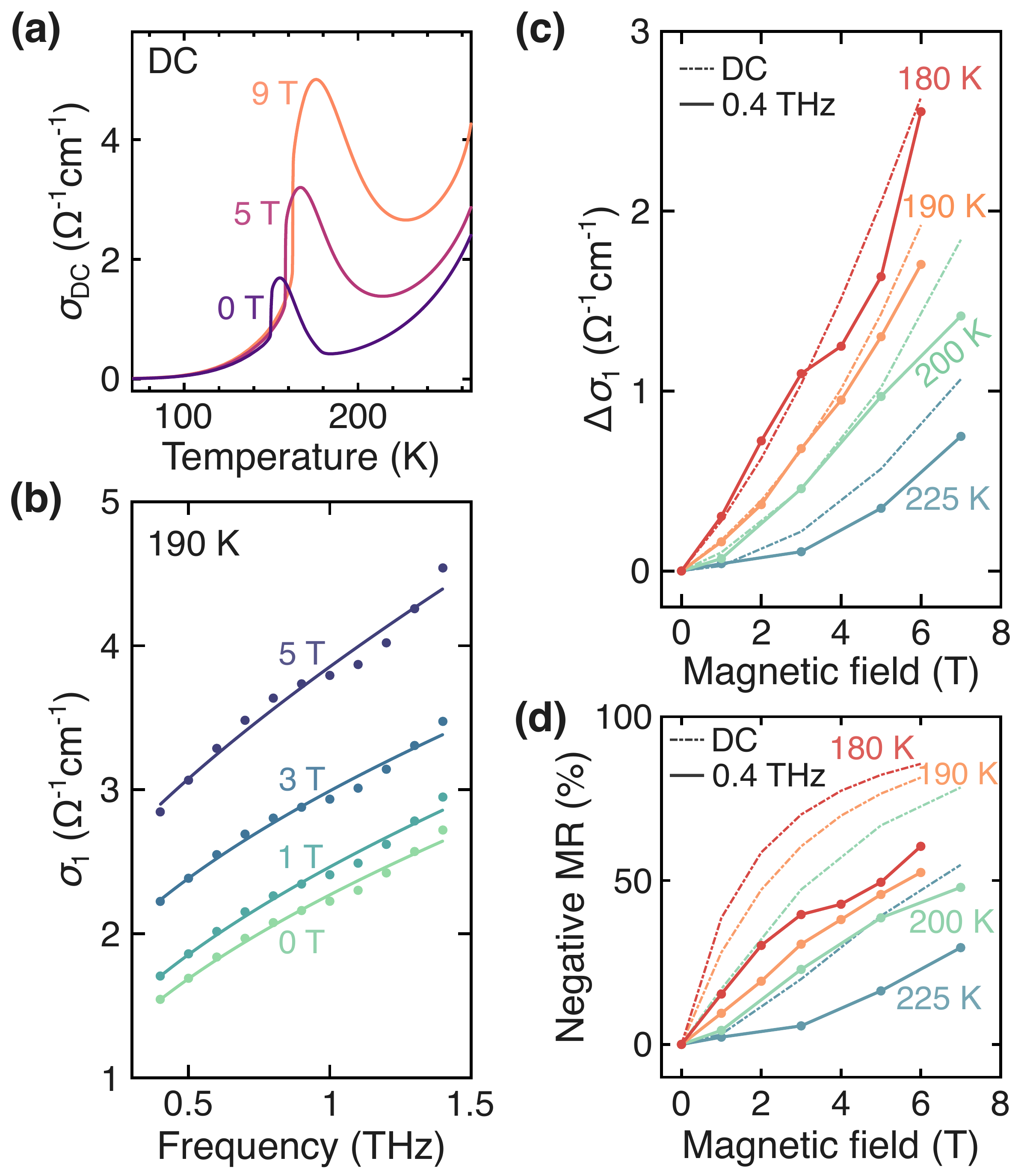}
    \caption{Colossal THz magnetoresistance observed in La$_{0.875}$Sr$_{0.125}$MnO$_3$. (a)~DC conductivity, $\sigma_\text{DC}$, versus $T$ at 0\,T, 5\,T, and 9\,T. (b)~Magnetic field dependence of $\sigma_1(\omega)$ at 190\,K. The circles represent experimental data, and the solid lines are power-law fits to the data. (c)~$H$-induced conductivity change, $\Delta\sigma_1$, at 0\,THz (DC) and 0.4\,THz at various $T$. (d)~Negative magnetoresistance at 0\,THz and 0.4\,THz at different $T$.}
    \label{fig:Bdepsigma}
\end{figure}

Figure~\ref{fig:time}(b) shows signals of the transmitted THz electric field as a function of time delay at various $T$. The $T$-dependence of the amplitude of the THz waveform is consistent with the phase diagram known for this compound~\cite{ZhouGoodenough2015PRB} and the DC conductivity [the lowest trace in Fig.~\ref{fig:Tdep}(a)]. In the highest $T$ range up to 250\,K, where the material is paramagnetic and insulating (PI), the amplitude increased with decreasing $T$. The sample is ferromagnetic and metallic (FM) between $T_\text{C}$ ($\sim$180\,K) and the charge ordering temperature, $T_\text{CO}$ ($\sim$150\,K). In this intermediate $T$ range, the amplitude of the time-domain traces decreased with decreasing $T$. Finally, a metal-insulator transition occurs at $T_\text{CO}$, below which the material is insulating and ferromagnetic (FI), where the amplitude increased with decreasing $T$. Furthermore, the peak position of the THz waveform shifted to a later moment when $T$ decreased, especially when the sample went through the FM phase, reflecting the changes in the real part of the refractive index of the sample. Figure~\ref{fig:time}(c) shows THz electric-field waveforms at various $H$ at a constant $T$, at which the DC conductivity exhibits CMR [see Fig.~\ref{fig:Bdepsigma}(a)]. These data show that the amplitude of the THz electric-field waveform decreases with increasing $H$. In addition, the peak position of the time-domain traces shifted gradually to a later moment with increasing $H$. 

We Fourier-transformed the THz electric-field waveforms, $E_{\text{sample}}(t)$ and $E_{\text{reference}}(t)$, at each $T$ and $H$, into complex-valued frequency-domain spectra, $\tilde{E}_{\text{sample}}(\omega)$ and $\tilde{E}_{\text{reference}}(\omega)$, respectively. By minimizing the deviations between the experimental and theoretical complex transfer function, $\tilde{E}_{\text{sample}}(\omega)/\tilde{E}_{\text{reference}}(\omega)$, one can obtain the complex refractive index of the sample, $\tilde{n}(\omega)$, without any approximation. We utilized open-source optimization software, Nelly~\cite{TayvahEtAl2021AC}. $\tilde{n}(\omega)$ can be converted into the complex conductivity, $\tilde{\sigma}(\omega) = \sigma_1(\omega) + i\sigma_2(\omega)$, and the complex permittivity (or dielectric constant), $\tilde{\varepsilon}(\omega) = \varepsilon_1(\omega) + i\varepsilon_2(\omega)$, through the relation
\begin{equation}
    \tilde{n}(\omega)^2 = \tilde{\varepsilon}(\omega) = \varepsilon_{\text{lattice}} + i\,\frac{\tilde{\sigma}(\omega)}{\omega\varepsilon_0},
\end{equation}
where $\varepsilon_0$ is the vacuum permittivity, $\varepsilon_{\text{lattice}}$ is a constant value to account for the increase in the permittivity caused by the lattice resonances that exist at high frequencies outside our spectral range. In this work, we focus on $\sigma_1(\omega)$ and $\varepsilon_1(\omega)$ at different $T$ and $H$ because they are the quantities that describe the MR and MC in the sample.

The $T$ dependence of $\sigma_1$ at four different frequencies, $\omega/2\pi =$ 0\,THz (DC), 0.4\,THz, 1.0\,THz, and 1.4\,THz, at $\mu_0H=0$\,T are plotted in Fig.~\ref{fig:Tdep}(a). No traces are intentionally offset in Fig.~\ref{fig:Tdep}(a). The $T$ dependence of $\sigma_1(\omega)$ in the THz frequency range showed agreement with the $T$ dependence of $\sigma_\text{DC}$, where $\sigma_1(\omega)$ increased (decreased) with decreasing $T$ in the metallic (insulating) phase. The phase transitions at $T_\text{CO}$ and $T_\text{C}$ are clearly revealed in the $T$ dependence of $\sigma_1(\omega)$. Note that $\sigma_1$ at 0.4\,THz and 1.0\,THz showed another insulator-to-metal-like transition at $\sim$20\,K by an upturn of conductivity with decreasing $T$; this behavior is in agreement with recent proposals for the coexistence of the FI and FM phases below $\sim$30\,K~\cite{PapavassiliouEtAl2006PRL,WeiEtAl2010NJP}. DC conductivity measurements on a different crystal with a small current at temperatures down to 30\,K (see Supplement 1) revealed a gap of 19.6\,meV (or 4.7\,THz) near 30\,K. The substantial gain of optical conductivity for frequencies below 4.7 THz in Fig.~\ref{fig:Tdep}(a) implies that carriers are released from the charge/orbital ordered state below $T_\text{CO}$ by absorbing THz photons. The photon-induced carriers may be responsible for the metallic-like conductivity below 30\,K for $\omega/2\pi=$ 0.4--1.0\,THz.

The phase transitions also manifested themselves in the $T$ dependence of $\varepsilon_1$, as shown in Fig.~\ref{fig:Tdep}(b). As $T$ is lowered from the PI-phase side, a rapid increase in $\varepsilon_1$ occurred as the system went through the FM phase, i.e., between $T_\text{C}$ and $T_\text{CO}$. Once the system entered the FI phase, however, the rapid change suddenly stopped and $\varepsilon_1$ increased only gradually with further decreasing $T$. The absolute (relative) permittivity change that occurs when the system passes through the FM phase, $\varepsilon_1(T_\text{CO}) - \varepsilon_1(T_\text{C})$, reached $\sim$7 ($\sim$26\%) at 0.4\,THz. The observed sharp increase of $\varepsilon_1$ with decreasing $T$ in the FM phase cannot be explained by the Drude model. 

Spectra of $\sigma_1(\omega)$ and $\varepsilon_1(\omega)$ measured at various $T$ are shown in Figs.~\ref{fig:Tdep}(c) and (d). The frequency dependence of $\sigma_1$ and $\varepsilon_1$ did not change greatly with decreasing $T$, except that the area under the curves changed. The real conductivity $\sigma_1(\omega)$ always exhibited a non-Drude behavior, i.e., $\sigma_1$ increases with increasing $\omega$, even in the FM phase, indicating that the ``metallic'' behavior in this phase cannot be described by a simple free electron model. The $\varepsilon_1$ spectra always exhibited a negative slope at all temperatures. Our results are consistent with the spectra extracted from previous measurements using a set of continuous-wave gigahertz sources in a Mach--Zehnder interferometer arrangement~\cite{PimenovEtAl1999PRB}. The $\sigma_1$ and $\varepsilon_1$ spectra can be fit well separately by the universal power law for amorphous and composite materials~\cite{Jonscher1977N,Jonscher1983,Elliott1987AP} (Supplement 1), as shown by the solid lines in Figs.~\ref{fig:Tdep}(c) and (d).

At $T\sim T_\text{C}$, CMR appeared in DC transport when an external magnetic field was applied, as shown in Fig.~\ref{fig:Bdepsigma}(a). The DC conductivity, $\sigma_{\text{DC}}$, sensitively increased with $H$ at $T>T_\text{C}$. $T_\text{CO}$ increased slightly with $H$. In addition, we also observed CMR in the THz frequency range, as shown in Fig.\,\ref{fig:Bdepsigma}(b), which plots $\sigma_1(\omega,H)$ versus $\omega$ for $T=190$\,K and $\mu_0H=$ 0, 1, 3, and 5\,T. The magnitude of $\sigma_1$ increased with $H$ in the THz range in a similar way to DC, i.e., there are no drastic spectral changes with $H$. Furthermore, Fig.~\ref{fig:Bdepsigma}(c) shows a conductivity change, $\Delta\sigma_1$, versus $H$ at $\omega/2\pi=0$\,THz (DC) and 0.4\,THz for four different $T$ near and above $T_\text{C}$. $\Delta\sigma_1$ was more significant when $T\sim T_\text{C}$, reaching a maximum at $\sim$180\,K. Notably, the magnitude of $\Delta\sigma_1(\omega/2\pi=\text{0.4\,THz})$ was comparable to $\Delta\sigma_1(\omega/2\pi=\text{0\,THz})$. We found that the $H$ dependence of $\Delta\sigma_1$ resembled a quadratic function at high $T$ but turned into a linear function gradually with decreasing $T$. Although the absolute magnitude of $\Delta\sigma_1$ at 0.4\,THz was close to that at 0\,THz, the value of the negative $\text{MR}= \left \{ \rho(\omega,\mu_0H=0\,\text{T})-\rho(\omega,\mu_0H) \right \}/\rho(\omega,\mu_0H=0\,\text{T})$, where $\rho(\omega,H)=1/\sigma_1(\omega,H)$, differed at those frequencies; see Fig.~\ref{fig:Bdepsigma}(d). This can be explained by the fact that $\rho$ at DC is much lower than that at 0.4\,THz when $\mu_0H=0$\,T. For instance, the negative MR reached 86\% (60\%) at $\omega/2\pi=0$\,THz (0.4\,THz), $T=180$\,K, and $\mu_0H=6$\,T.

\begin{figure}[hbt!]
    \centering
    \includegraphics[width=0.45\textwidth]{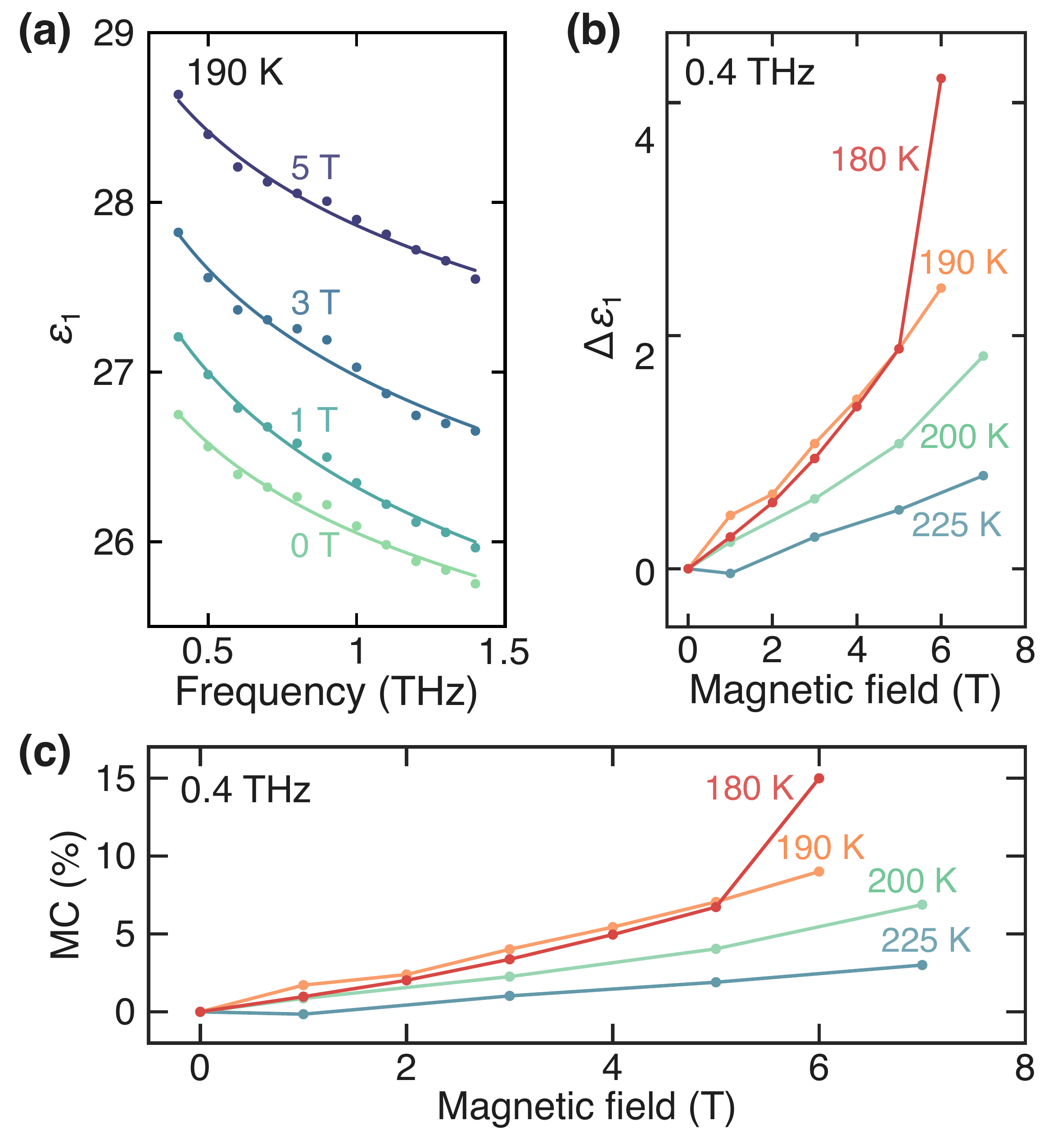}
    \caption{Colossal THz magnetocapacitance in La$_{0.875}$Sr$_{0.125}$MnO$_3$. (a)~$H$ dependence of $\varepsilon_1(\omega)$ at 190\,K. The circles represent experimental data, and the dashed lines are power-law fits to the data. (b)~$H$-induced permittivity change, $\Delta\varepsilon_1$, at 0.4\,THz at different $T$. (c)~Magnetocapacitance at 0.4\,THz at different $T$.}
    \label{fig:Bdepeps}
\end{figure}

Furthermore, we found that the magnitude of $\varepsilon_1$ increased considerably in response to an external magnetic field when $T\sim T_\text{C}$ while the spectral shape of $\varepsilon_1(\omega)$ did not change significantly with $H$; see Fig.~\ref{fig:Bdepeps}(a). Figure~\ref{fig:Bdepeps}(b) shows the permittivity change, $\Delta\varepsilon_1(\omega,H)$, as a function of $H$ at 0.4\,THz. The value of $\Delta\varepsilon_1(\omega,H)$ was greatest when $T\sim T_\text{C}$ in a similar manner to $\Delta\sigma_1(\omega,H)$. 
The highest value of $\Delta\varepsilon_1(\omega/2\pi=0.4\,\text{THz},\mu_0H)$ in our measurements was 4.2 at 180\,K at 6\,T, corresponding to 15\% in $\text{MC}=\Delta\varepsilon_1(\omega,\mu_0H)/\varepsilon_1(\omega,\mu_0H=0\,\text{T})$; see Fig.~\ref{fig:Bdepeps}(c). Note that the sharp jump in $\varepsilon_1$ within the FM phase with decreasing $T$ has been previously observed, and it has been attributed to a polaron-ordering transition~\cite{IvanovEtAl1998JAP} and a result from the onset of magnetic ordering~\cite{PimenovEtAl1999PRB}. The observed $H$-induced increase in $\varepsilon_1$ in our data suggests that this phenomenon is closely related to the spin ordering near $T_\text{C}$.

The observed phenomenon can be qualitatively explained by a two-component model consisting of two phases with different $\sigma_1$~\cite{EfrosShklovskii1976PSSB,BergmanImry1977PRL}. Here we use P$_\text{L}$ and P$_\text{H}$ to denote the phases with a lower and higher $\sigma_1$, respectively. Let the fraction of P$_\text{H}$ be $p$, i.e., the system is completely in P$_\text{L}$ (P$_\text{H}$) when $p=0$ ($p=1$). Percolation theory suggests that the $\sigma_1$ of the system increases with increasing $p$, as shown below\cite{EfrosShklovskii1976PSSB}:
\begin{equation}
    p>p_\text{c}: \sigma_1 = \sigma_\text{H}\left(p-p_\text{c}\right)^t,
\end{equation}
\begin{equation}
    p<p_\text{c}: \sigma_1 = \sigma_\text{L}\left(p_\text{c}-p\right)^{-q},
    \label{eqn:percolationsigma}
\end{equation}
where $t$ and $q$ are critical indices. Furthermore, percolation theory suggests that $\varepsilon_1$ increases sharply with increasing $p$ when $p$ is close to but lower than the percolation threshold, $p_\text{c}$~\cite{EfrosShklovskii1976PSSB}. The increase in $\varepsilon_1$ can be understood as the increase in capacitance by decreasing the separation between two metal plates~\cite{DubrovEtAl1976SJETP,EfrosShklovskii1976PSSB}. Therefore, both $\sigma_1$ and $\varepsilon_1$ increase with increasing $p$ in the model. 

For mixed-valence manganites, the presence of inhomogeneities has been reported and extensively discussed~\cite{MoreoEtAl1999S,DagottoEtAl2001PR}. Electron transport in the P$_\text{H}$ phase in La$_{0.875}$Sr$_{0.125}$MnO$_3$ is believed to be governed by the double exchange interaction~\cite{Zener1951PR}, which suggests that electrons can hop more easily when neighboring spins are parallel. Hence, $p$ should be determined by the degree of magnetic ordering and thus increase with either decreasing $T$ or increasing $H$, leading to the simultaneous increase of $\sigma_1(\omega)$ and $\varepsilon_1(\omega)$ near $T_\text{C}$. The fraction $p$ continuously increases with decreasing $T$ until $T$ reaches $T_\text{CO}$, where the two-component model is expected to break down due to the emergence of charge and orbital ordering and the first-order structural phase transition. It has been proposed that an orbital polaron lattice forms in the FI phase below $T_\text{CO}$, suppressing the hopping of charge carriers~\cite{GeckEtAl2005PRL}. Due to the complexity of the FI phase, a discussion of the magnetic field responses in the FI phase is deferred to future work. Moreover, our observation that $\varepsilon_1(\omega)$ monotonically increases with decreasing $T$ for $T>T_\text{CO}$ suggests that $p$ is still below $p_\text{c}$ when $T=T_\text{CO}$.

It is expected that $p$ increases upon strontium doping because the doping increases the concentration of Mn$^{4+}$ sites. The charge transfer over Mn$^{3+}$--O--Mn$^{4+}$ arrays facilitates the double exchange interaction. It has been shown that the sharp increase in $\varepsilon_1$ with decreasing $T$ reaches its maximum at $x\sim0.16$~\cite{PimenovEtAl1999PRB}. This may imply that $p$ is close to $p_\text{c}$ for the two-component model when $x\sim0.16$. Interestingly, $x\sim0.16$ is also the percolation threshold for the Griffiths phase (the coexistence of paramagnetic and ferromagnetic clusters) above $T_\text{C}$~\cite{DeisenhoferEtAl2005PRL} and the critical concentration for the metal-insulator transition (the FI phase at $T$ below $T_\text{CO}$ disappears when $x\gtrsim0.16$)~\cite{ZhouGoodenough2015PRB}. This suggests that the FI phase at low temperatures, the Griffiths phase, and the sharp increase in $\varepsilon_1(\omega)$ with decreasing $T$ or increasing $H$, are closely related. 

\section{Conclusion}
In summary, we have demonstrated the coexistence of colossal THz MR and MC in a bulk single crystal of La$_{0.875}$Sr$_{0.125}$MnO$_3$ at $T\sim T_\text{C}$. The colossal THz MR and MC did not show saturation with increasing magnetic field up to 6\,T. The $\sigma_1$ and $\varepsilon_1$ spectra showed opposite trends to the Drude model at all temperatures below 250\,K. The spectral shapes are not sensitive to the applied magnetic field. Further work needs to be performed to establish whether the coexistence of colossal THz MR and MC is a generic feature of perovskite manganites that exhibit CMR, which might provide new insights into the origin of CMR. We showed that the simultaneous increase of the $\sigma_1$ and $\varepsilon_1$ can be explained by a two-component model in percolation theory. Note that the conductivity of La$_{0.875}$Sr$_{0.125}$MnO$_3$ is relatively low compared to metallic systems (weak THz absorption) and no sharp features were observed in the spectra. This suggests that La$_{0.875}$Sr$_{0.125}$MnO$_3$ is suitable for THz devices that can utilize the large magnetic amplitude and phase modulations of THz waves. The modulation depth and the operating temperature can be increased by increasing the doping concentration ($T_\text{C}\sim288\,K$ for $x=0.175$~\cite{PimenovEtAl1999PRB}) and employing nanostructure designs~\cite{Lloyd-HughesEtAl2017NL}.

\begin{acknowledgments}
This research was primarily supported by the National Science Foundation through the Center for Dynamics and Control of Materials: an NSF MRSEC under Cooperative Agreement No. DMR-1720595. G.A.F. acknowledges additional support from NSF Grant DMR-2114825. The authors wish to thank X. Li for performing some preliminary measurements.
\end{acknowledgments}


\bibliography{ms}

\end{document}


\title{Supplementary Material for Observation of Colossal Terahertz Magnetoresistance and Magnetocapacitance in a Perovskite Manganite}

\author{Fuyang~Tay}
\author{Swati~Chaudhary}
\author{Jiaming~He}
\author{Nicolas~Marquez~Peraca}
\author{Andrey~Baydin}
\author{Gregory~A.~Fiete}
\author{Jianshi~Zhou}
\author{Junichiro~Kono}

\maketitle

\section{DC resistivity measurement}
\begin{figure}[htbp]
\centering\includegraphics[width=0.4\textwidth]{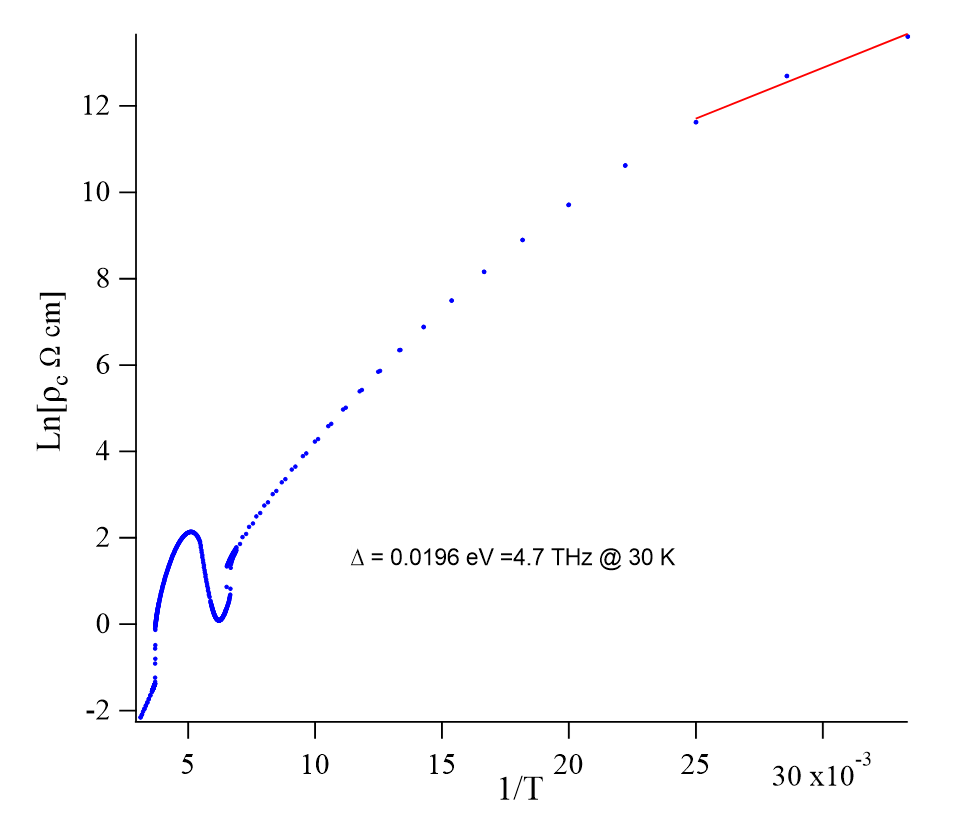}
\caption{Temperature dependence of DC resistivity for a different La$_{0.875}$Sr$_{0.125}$MnO$_3$ crystal with a small current.}
\label{fig:DCrhovsT}
\end{figure}

The temperature dependence of DC resistivity in Fig.~\ref{fig:DCrhovsT} indicated a gap of 19.6\,meV (or 4.7\,THz) near 30\,K.

\section{Polarization dependence of terahertz transmission}
\label{sec:anisotropy}
\begin{figure}[htbp]
\centering\includegraphics[width=0.25\textwidth]{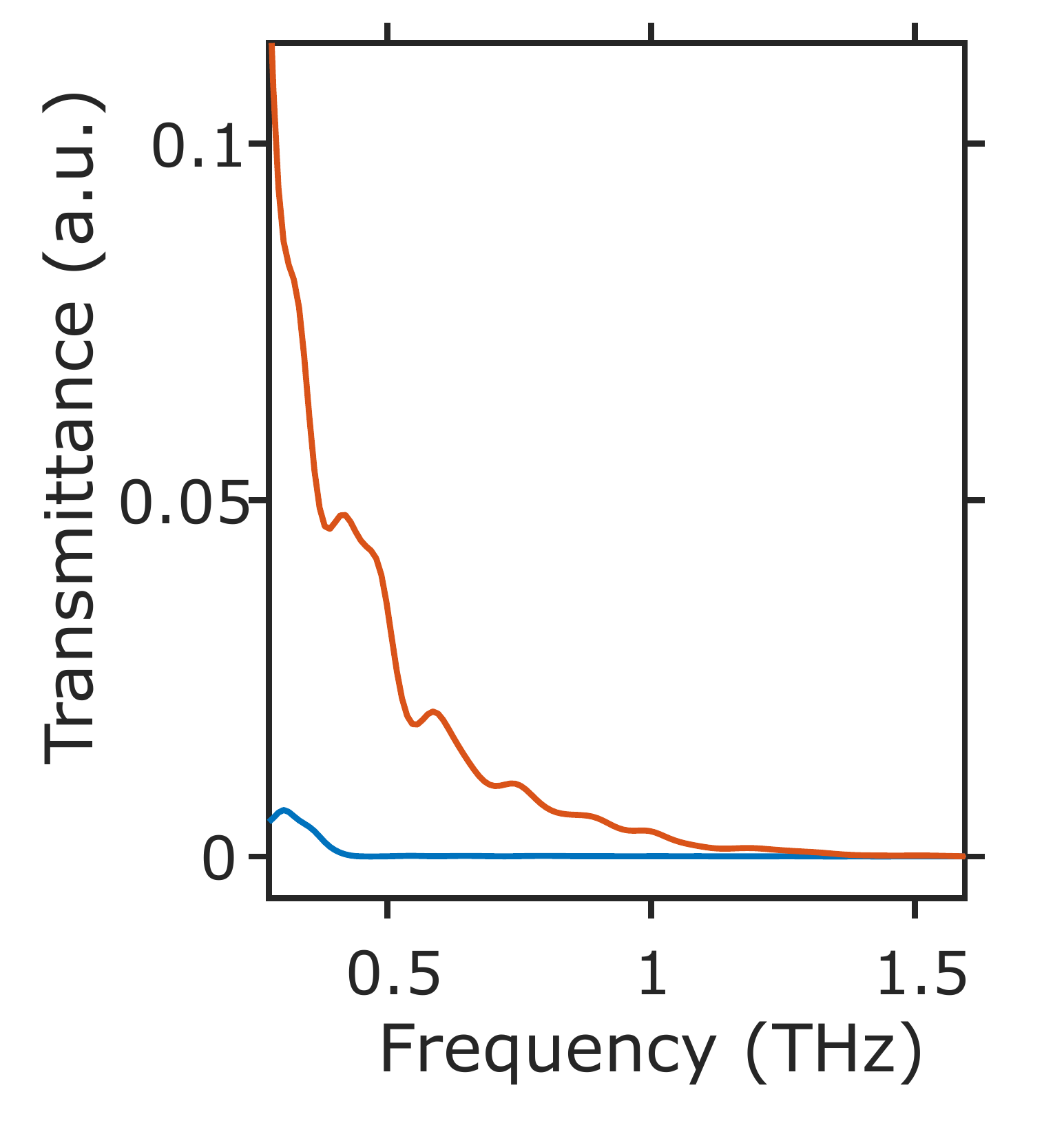}
\caption{Transmittance spectra when the THz electric field was polarized along the $a$ and $b$ axes, represented by red and blue lines, respectively.}
\label{fig:anisotropy}
\end{figure}

We investigated THz transmittance spectra when the THz electric field polarization was parallel to the $a$ and $b$ axes. Figure~\ref{fig:anisotropy} shows transmittance spectra at 4.2\,K. The transmittance of a THz wave with the electric field polarized along the $a$ axis was much higher than that of a THz wave polarized along the $b$ axis. The anisotropy existed at all temperatures below 250\,K. Such behavior has been previously reported~\cite{PimenovEtAl1999PRB}.

\section{Power-law fittings}
\label{sec:powerlaw}
The frequency dependence of the complex permittivity (or dielectric constant) of dielectric systems, in particular amorphous and composite materials, have been found to follow a universal behavior~\cite{Jonscher1977N,Jonscher1999JPDAP}. In general, one can express $\sigma_1(\omega)$ and $\varepsilon_1(\omega)$ as
\begin{align}
    \sigma_1(\omega) &= \sigma_{\text{DC}} + A\omega^s,\label{eqn:powerlaw_sigma}\\
    \varepsilon_1(\omega) &= A\omega^{s-1}\varepsilon_0^{-1}\tan{\left (\frac{s\pi}{2} \right )},
    \label{eqn:powerlaw_eps}
\end{align}
where $\sigma_{\text{DC}}$ is the DC conductivity, $\varepsilon_0$ is the vacuum permittivity, and $A$ and $s$ are fitting parameters for both $\sigma_1(\omega)$ and $\varepsilon_1(\omega)$. However, we found that the $\sigma_1(\omega)$ and $\varepsilon_1(\omega)$ obtained from our experiments cannot be fit well simultaneously by Eqns.~(\ref{eqn:powerlaw_sigma}-\ref{eqn:powerlaw_eps}). Instead, good agreement was obtained when $\sigma_1(\omega)$ and $\varepsilon_1(\omega)$ were fit individually, i.e., different sets of optimum values of $A$ and $s$ were obtained for $\sigma_1$ and $\varepsilon_1$ spectra. This may be caused by other mechanisms that are not considered in the dielectric relaxation model, such as the contribution from phonons~\cite{PimenovEtAl1999PRB}. Furthermore, the $\sigma_\text{DC}$ obtained from the DC measurements did not account for the anisotropy discussed in Section~\ref{sec:anisotropy}.

\bibliography{supplement}